# Helium: Visualization of Large Scale Plant Pedigrees

Paul D Shaw[1,2]*, Martin Graham[2], Jessie Kennedy[2], Iain Milne[1] and David F Marshall[1]

**Abstract**

**Background:** Plant breeders are utilising an increasingly diverse range of data types in order to identify lines that have desirable characteristics which are suitable to be taken forward in plant breeding programmes. There are a number of key morphological and physiological traits such as disease resistance and yield that are required to be maintained, and improved upon if a commercial variety is to be successful. Computational tools that provide the ability to pull this data together, and integrate with pedigree structure, will enable breeders to make better decisions on which plant lines are used in crossings to meet both critical demands for increased yield/production and adaptation to climate change.

**Results:** We have used a large and unique set of experimental barley (*H. vulgare*) data to develop a prototype pedigree visualization system and performed a subjective user evaluation with domain experts to guide and direct the development of an interactive pedigree visualization tool which we have called Helium.

**Conclusions:** We show that Helium allows users to easily integrate a number of data types along with large plant pedigrees to offer an integrated environment in which they can explore pedigree data. We have also verified that users were happy with the abstract representation of pedigrees that we have used in our visualization tool.

## Background

The effects of climate change and ensuring food security in a world with an increasing population is becoming ever more pertinent [1, 2, 3]. The exploitation of pedigrees in plant breeding allows breeders to target specific plant crosses to maximise the potential of achieving desirable agriculturally important characteristics such as yield, drought/water tolerance and disease resistance which will be required if new varieties are to be bred to cope with increased demand in a changing environment.

The ability to predict and visualize the inheritance of alleles that facilitate resistance to pathogens or any other commercially important characteristic is crucially important to experimental plant genetics and commercial plant breeding programmes. Derivation of the inheritance of such traits by traditional molecular techniques is expensive and time consuming, even with recent developments in high-throughput technologies. This is especially true in industrial settings where, due to time constraints relating to growing seasons, many thousands of plant lines may need to be screened quickly, efficiently and economically every year.

Due to their complexity, there is a cognitive limitation in conceptualising large pedigree structures.

While it may not be achievable or indeed necessary to understand every mating relationship between related individuals, an overall picture can lead to insight into the data and any patterns it may contain. This can also aid in the identification of problems (both biological and data handling issues) within datasets when coupled with expert domain knowledge.

This is particularly important when looking at pedigree data as the context in which each line sits may hold additional and important information (such as the inheritance of particular genome regions from ancestral varieties). It is because of this that a combination of visual and statistical analytics would allow geneticists and commercial breeders to gain a deeper understanding of the transmission of genetic elements within a pedigree based framework but there is currently a lack of suitable tools to analyse these data types.

*Correspondence: paul.shaw@hutton.ac.uk
[1]Information and Computational Sciences, The James Hutton Institute, Invergowrie, DD2 5DA Dundee, UK
[2]School of Computing, Edinburgh Napier University, 10 Colinton Road, EH10 5DT Edinburgh, UK
Full list of author information is available at the end of the article
†Equal contributor

Software tools that offered improvements in the speed at which this analysis can be carried out, and increase users' ability to conceptualise large pedigrees would bring both time and cost gains to breeding companies.

Using a unique and extensive barley dataset covering pedigree, genotypic and phenotypic data for UK elite germplasm which has been through the UK National List Testing procedures [4], we discuss the challenges of visualizing the transmission of alleles encoding traits and characteristics of agricultural importance in a pedigree-based framework. We then describe the subsequent development of a pedigree visualization tool which we have implemented in close collaboration with domain experts.

## Plant pedigrees

A pedigree (Figure 1) is a representation of how genetically discrete individuals are related (usually but not exclusively) in time to one another. It is therefore a representation of the genetic relationship between individual plant lines, their parents and progeny (predecessors and successors). Pedigrees are often used in human contexts to show the transmission of alleles responsible for genetic conditions of medical importance. In plants they are used as a framework along with environmental data, on which statistical analysis can be used to determine factors such as mode of inheritance (Identity by Descent, IBD and Identity by Association, IBA). Additionally, they are often used to check for potential genotyping errors, since these errors, by the very nature of Mendelian inheritance, are constrained by the pedigree structure in which they exist [5]. The accurate representation of pedigrees is therefore becoming increasingly important in plant breeding and genetics.

While there are defined standard nomenclatures for human pedigrees [6] there is no *single* formal system for plant pedigrees, however, there are moves towards defining standards. There are valid biological reasons for this including: the hermaphrodite nature of most plant species, the complexity of mating designs possible in plant genetics and, finally, the absence of any overseeing coordinating organisation.

While plant and animal breeding share routine breeding techniques such as standard crossing and back-crossing, pedigrees used in plant breeding display some subtle but important differences, often involving key shorthand conventions that are unique to plant mating designs leading to complex textual based records which can be difficult to read (Figure

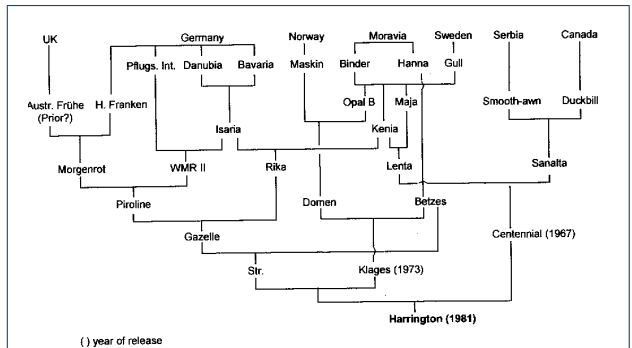

**Figure 1 Traditional barley pedigree.** Common representation of a barley pedigree showing Elite cultivars [7]. These representations cover only a limited number of lines, are commonly seen in humans and animals and are therefore easy to read.

2). Firstly, the named entities in plant pedigrees may, but not always, represent a population of genetically identical individuals, not a single plant. While it is relatively simple to grow many plants from seed, potentially many decades after production, in humans and animals this is understandably not the norm. The generation of these genetically identical (homozygous) varieties is possible through doubled haploidy, inbreeding, or crossing of pairs of inbred lines to achieve what is termed an F1 hybrid. Successive inbreeding by self-pollination of these F1 generation plants leads to individual plants that are close to homozygous across all alleles. The exploitation of homozygous lines in crop species such as barley is a powerful tool in genetic analysis, removing some of the genetic complexities associated with species (such as humans) where there is a high level of heterozygosity.

a. A/B//C//D
b. ((A * B) * C) * D
c. [ A x [(B x C) * D] x E] * [F x A] x C

**Figure 2 Example pedigree formats.** Pedigree formats can be complex with no standard nomenclature; a. Purdy Notation System [8] was put forward by Purdy as a common format for representing small grain cereal pedigrees. Forward slashes '/' are used to delimit lines. In this case A is crossed with B which is then crossed with C whose progeny is crossed with D. b. Lamacraft and Finlay notation [9] which was put forward as a format which could be more easily parsed by computers. The example here is the same as in the Purdy notation above. c. A typical pedigree that we find in old records where a mixture of notations are used. These mixed notation systems are common and most breeders will use shorthand that is unique to them. These records are sometimes difficult to read and would benefit from being represented in a more user friendly way.

## Genetic transmission

In genetic recombination the genetic composition of an individual is split, on average, 0.5: 0.5 between its parents if the result of cross fertilisation or 1 if the parent is self-fertilised (selfed). Non-parental alleles can only result from either a misclassified genotype or the result of genetic mutation. Alleles must therefore be inherited from either parent. If this is not the case then it suggests there are problems with either the underlying genotypic data or misclassification of plant lines.

## Data sets

There are a number of different data types that we have used in this work. Our primary data set is composed of a large barley pedigree data set for 803 UK Elite cultivars as well as Single Nucleotide Polymorphism (SNP) genotypic data for 750 of these lines across 4,769 genetic markers. In addition, we have phenotypic data for these lines for 33 Distinctiveness, Uniformity and Stability (DUS) characters [10] across multiple years and sites (1980 – present which equates to 601,148 data points). We also have datasets covering UK wheat (*Tritticum* spp.) and Asian rice (*Oryza sativa*) which we have used in our work although these are more limited in size. Data are stored in the Germinate 2 database system. [11]. The ability to connect to Germinate was an important design decision as we wanted to allow users to access all the background information on plant lines that we had available.

## Pedigree definitions

The nucleus of pedigree data are a series of parent/child relationships defined as encoded strings (Figure 2) [8, 9]. We have atomised our data into simple parent/child definitions which can be used to dynamically reconstruct the pedigree. In addition there may also be information identifying whether the parent was male or female and the type of genetic cross performed. Something unique in plant breeding is where a plant can be both male and female parents in the same cross.

Complications may arise from either older pedigree data which is error prone and may be difficult to verify without expert guidance and from the re-use of names to describe varieties creating false relationship joins. It is not uncommon for a breeder's favourite name to be used multiple times until a line is adequately different, and has sufficient performance to be accepted for wider distribution into the UK recommended list programme.

## Genotypic data

The genotypic data set for our study is based on a set of SNP markers which are mapped to known chromosome positions in the barley genome. Each plant line within the test set has been genotyped for a set of 7,000 of these markers.

A given plant variety will have an allele call for each of a series of loci represented as a pair of nucleotide bases e.g. AA, GG (which are homozygous) or AG (which are heterozygous), for a locus. Due to the inbred nature of our barley germplasm there are low levels (less than 0.5%) of residual heterozygosity present.

## Phenotypic data

The phenotypic data in our study has been either collected in field experiments or by molecular testing. Though many of the agriculturally important traits are controlled by many genes of small effect (quantitative traits) for simplicity we are concentrating on traits under simple genetic control. Examples of such traits include DUS characteristics which are used in the varietal registration and seed certification process and allele data on disease resistance genes such as *Mlo* and *Mla*.

# Previous work

The ability to visualise data is imperative in modern experimental plant genetics, with volumes of data being routinely produced far exceeding the ability for humans to digest and identify underlying phenomena. Until now, pedigree visualization, with few exceptions [12, 13] has primarily been focussed on work carried out in the human genetics domain. Because plant breeding programmes involve phenomena not normally seen in human populations, such as routine inbreeding, there are additional visualization challenges that need to be overcome. There are often large numbers of plant lines involved in any pedigree, many more so than in an average human pedigree due to factors such as generation time/time to sexual maturity which is far lower in most plant species than that of their mammalian counterparts. This section will look at the various visualization techniques used to represent pedigree based data and highlight the problems and strengths that these techniques exhibit.

## Table-based approaches

Table-based visualization tools such as Flapjack [14] address some of the problems associated with visualizing large datasets and are optimized for efficient sorting and querying of genotypic and phenotypic data, but currently lack the ability to display data on a pedigree-based scaffold.

While other tools such as PedStats [15] offer statistical validation of users' pedigree data without visualization of the actual pedigree structure, it is difficult if not impossible to conceptualize pedigree structure for complex data sets without some visual representation.

Matrix-based visualizations to represent pedigrees use the intersection of the x and y edge to define relationships. Matrix-based visualizations have advantages over node-link or graph-centred layout approaches including the ability to create compact graph representations and the ability to remove edge overlapping. However, tests generating matrix visualizations using our pedigree data have shown that the data density is so low the resulting representations are not particularly insightful. The ability to easily track flow and identify paths is also removed.

Tools such as GeneaQuilts [16], offer a new visualization technique suitable for use with thousands of individuals but offer limited scope for addition of complex genotypic and phenotypic data and discussions with our users showed that they found it difficult to easily interpret such representations.

Finally, tools such as VIPER [17] offer novel pedigree visualization and genotypic error checking capabilities. VIPER is essentially a stack of nested table representations of generations where rows represent sires, dams or children and columns represent individuals which can span multiple columns where they are parents. VIPER's primary use is in identification of genotyping problems in farmed animals and would be unsuitable for visualizing the complex crossing relationships that exist between crops where selfing is not uncommon. VIPER requires both separate male and female parents which is the norm in any applications handling animal or human data, but not always the case in plant breeding.

#### Graph-based

Unlike trees, graphs allow for the precise modelling of the complexity of a plant breeding programme. Techniques such as node link diagrams have long been used as a way of representing graph-based data and recent work has examined how effective the node-link model performs representing graph data when compared to matrix-based visualizations [18]. Work carried out by Purchase [19, 20] and Bennett [21] also indicated that while graph layout played an important part in a user's understanding, it was not the major focus; this focus perhaps being the use of other aesthetics relating to node colour and shape.

Most of the current tools have been developed for human pedigres where consanguineous mating events are negligible. This is not the case in plant and animal breeding which cannot be properly modelled using tools that use node-link or tree hierarchies such as Pedfiddler and Madeline [22].

Cranefoot [23] reports the use of mathematical graph structures to deal with between-relative mating but the approach is limited in its current form in the amount of information that can be attached to a node. Finally, HaploPainter [24] allows the drawing of genetic haplotypes, but suffers from being restricted in the number of individuals it is able to display.

A commonly used two-dimensional pedigree visualization tool is Peditree [13] which offers a tree-based view of data in a pedigree but this is not suited to our requirements as plant pedigrees are not trees (inbreeding and the use of older lines in more modern crosses prevents us from treating them as such). Other tools such as the Pedigree Visualizer by Wong [25] offer new layout algorithms. Wong suggests introducing duplicate "alias" lines in representations with multiple matings from the same individuals, phenomena that are commonplace in plant data. PyPedal [26] not only offers rudimentary graph drawing tools, restricted to changing node shape to represent male and females, but also error checking algorithms to try and identify potential pedigree errors where appropriate genotypic data exists.

Visualization techniques such as sunbursts [27] which are space filling versions of a node-link diagram have the advantage that a node's position in a hierarchy is maintained. Additionally, Fan Charts [28] and H-trees [29] have also been described as a means for recounting human genealogy; these techniques however assume no inbreeding (they are trees and not graphs) and thus rule themselves out for use with plant pedigrees.

While the main problems with these additional techniques are that they are not appropriate for observing a pedigree in its entirety (indeed the complexity of the data may rule many of them out), they may be useful when trying to visualize a sub-section of data such as a sub-pedigree for specific lines.

#### Layout algorithms

Plant pedigrees often form what we describe as a *pedigree net*, whereby there is structure to the graph but it's not as simple as traditional top-down pedigree representation that is seen in humans and to a lesser extent in farmed animals (Figure 3).

This abstract representation does include a time component in the form of generations, but due to the viability of seed, and the existence of varieties and landraces that may be many hundreds of years old, there is the potential to use these older varieties in modern crosses. This situation leads to nodes at the top of the graph having edges connecting to nodes at the bottom - this is not common in animals and would be extremely unlikely in humans. The existence of a time component means that the use of a layout algorithm that preserves topology (top-down generations) is nonetheless important as most (but not all) crossing will be between newer varieties. Because of this, layout methodologies such as force-directed algorithms (Figure 4B) would not offer the ability for us to arrange our pedigree based on time. Force directed layouts are not well suited to our requirements.The lack of a visually identifiable pedigree structure is strikingly apparent.

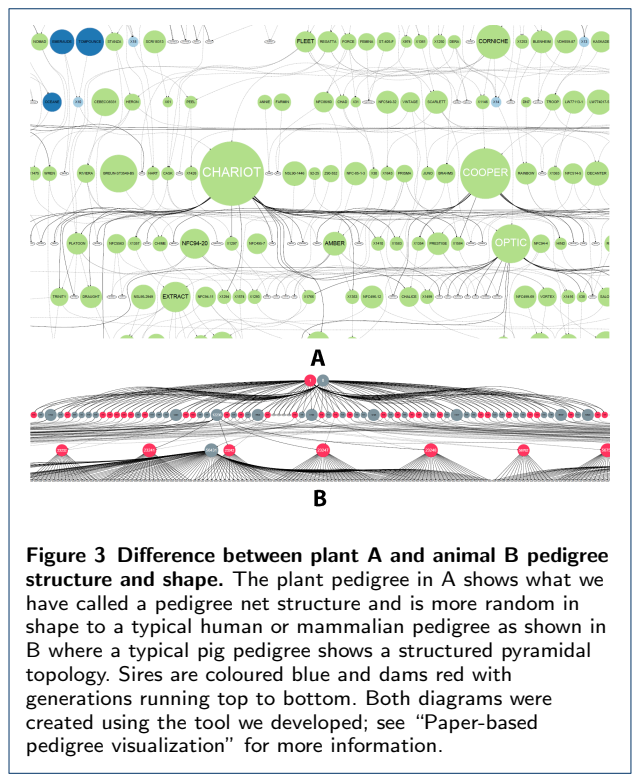

**Figure 3 Difference between plant A and animal B pedigree structure and shape.** The plant pedigree in A shows what we have called a pedigree net structure and is more random in shape to a typical human or mammalian pedigree as shown in B where a typical pig pedigree shows a structured pyramidal topology. Sires are coloured blue and dams red with generations running top to bottom. Both diagrams were created using the tool we developed; see "Paper-based pedigree visualization" for more information.

The problem of very large pedigrees in humans has been identified and solutions proposed in tools such as PViN [30] which looks at windows on large datasets but only offers pedigree drawing with no scope for addition of other information onto the visualization. In addition, its traditional human family tree output is not the most efficient use of space for plant pedigrees which form a more dense net due to the nature of reproduction which is not seen in humans or animals (Figure 3A).

Although there are problems associated with 2D node-link layouts such as a lack of horizontal space and problems with crossing of edges [31] they are still well suited to displaying data of this type. 3D tools also have their problems, including visual occlusion and that they tend to visualise high-level features and not specifics, so while some trends are easy to spot, the actual detail is hidden from the user. From this point of view they are limited in use for our purposes and offer no advantages over their 2D counterparts. Notable examples of such tools are Walrus [32] and Celestial3D [31] but their success lies in alternate problem domains.

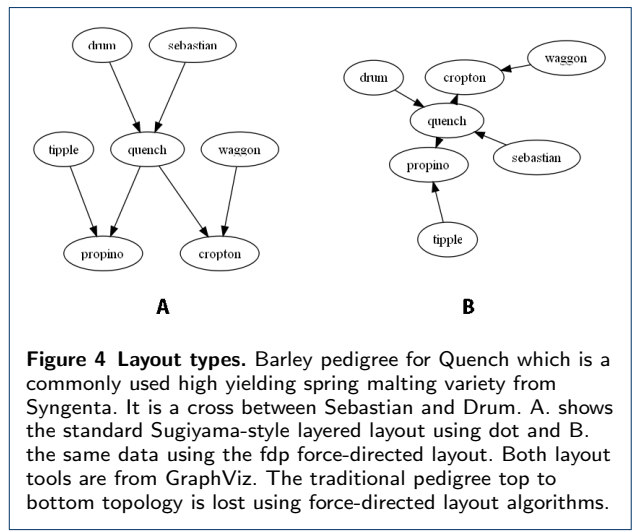

**Figure 4 Layout types.** Barley pedigree for Quench which is a commonly used high yielding spring malting variety from Syngenta. It is a cross between Sebastian and Drum. A. shows the standard Sugiyama-style layered layout using dot and B. the same data using the fdp force-directed layout. Both layout tools are from GraphViz. The traditional pedigree top to bottom topology is lost using force-directed layout algorithms.

## Discussion

It is clear that the techniques and tools that we have examined contain many features that are useful, but none meet the exact requirements (including data abstraction) of our problem to be able to overlay genotypic and phenotypic data onto a complex pedigree structure.

There is a need for the development of tools that are tailored for the unique needs of plant breeding with the ability to explore pedigree structure, and paint additional genotypic and phenotypic data on top, to allow breeders to make informed decisions and visualize the way in which alleles for agriculturally important traits are transmitted through previous and subsequent generations. Such tools do not currently exist.

Through the examination of methodologies to display pedigree data we suggest that the best method to visualize plant pedigree data is a layered layout (Sugiyama-style) based approach (Figure 3A and 4A). Not only does this allow us to accurately map the exact specifics of how breeding programmes run (including inbreeding) but also provides a well established framework onto which we can build our visualization. The use of graphs as our data structure means that features such as standard graph-traversal algorithms can be used to bring greater functionality to our pedigree structure in locating ancestors and descendants and as a logical framework on to which we can look for problems with our underlying datasets. The layered layout representation also brings a coherent structure to sparse relationships when compared to matrix style visualizations and generations and topological layout are more clear. This is not the case with animal (Figure 3B) and human pedigrees whose top-down fan type shape is not well suited to a layered layout as they quickly become very large, consuming large volumes of horizontal space [17].

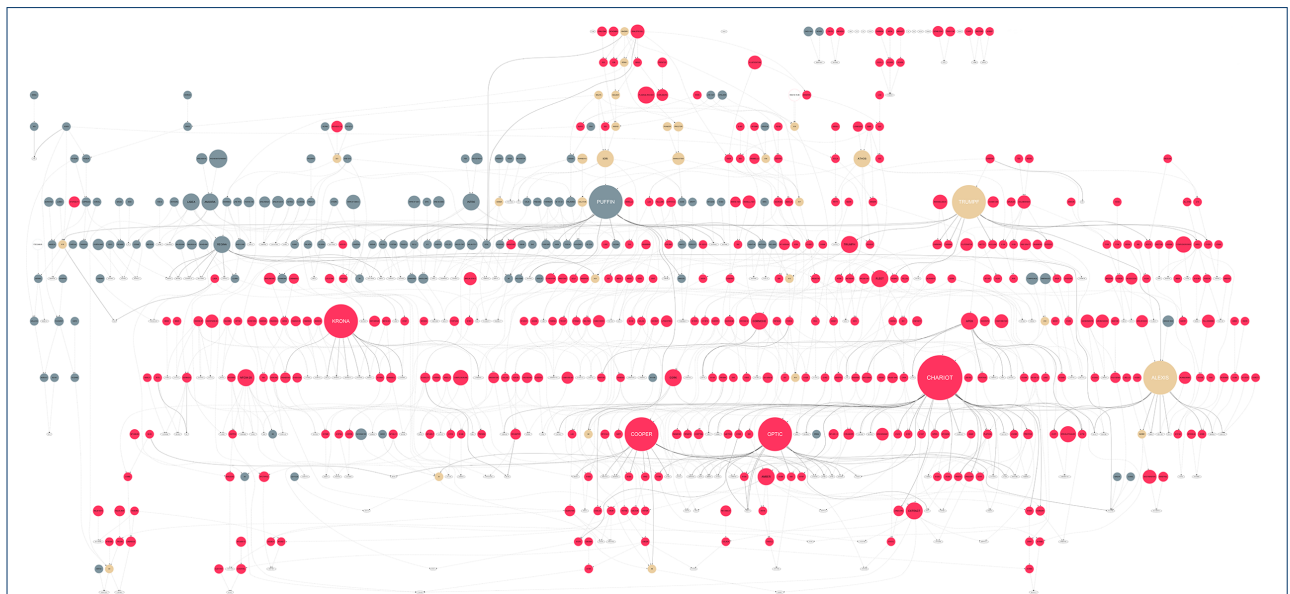

**Figure 5 Pedigree visualization static prototype.** This was one of our first attempts at visualizing our entire barley pedigree. The colours of nodes were used to distinguish between the winter/spring ecotype (red shows spring barley, blue shows winter barley and the cream coloured nodes are lines that are in both winter and spring pedigrees - qualitative data type) and node size to show the number of times the line has been used in crosses that have given rise to progeny that have been successful in National List trialling in the UK - quantitative data type. To the best of our knowledge this is the first time that a pedigree involving this number of commercially released lines has been brought together in one place and sparked interest with commercial plant breeders when they were presented with it.

Tools that allow exploration of data to try and bring a greater understanding of complex relationships between individuals should bring greater insight into how plant breeding programmes operate at the genetic level and how to bring maximum potential benefit from them. The ability to detect patterns and associations (or even anomalies) within these datasets such as; the identification of problems with inheritance of alleles, the identification of lines from which additional information would allow inference of data on large parts of the pedigree, simple typos and errors, or looking for lines which are similar to unknown lines, will lead to increased depth of domain knowledge for plant breeders and geneticists.

## Paper-based visualization

We wanted to test if our use of a DAG based data structure and layered layout approach would work with our barley pedigree data and would be accepted by our users. In order to do this we implemented a paper based layout overlaying basic character data on to the graph nodes represented by colour and sizing nodes based on the number of times they had been used in crosses in our data. In this prototype (which we implemented in Perl and the Graphviz dot library) we modelled our pedigree as graph nodes to represent plant lines and edges to show mating/parentage. While

GraphViz has been used before in pedigree drawing [33], examples focus on a small number of individuals.

While initially this prototype was run by users as a command-line computer program which generated images based on input files and generated an image which could be viewed on their computer monitors we decided that printing this static representation (2.5m x 1m see Figure 6) would allow domain experts to better interact with the visualization. We overlaid, by means of colouring nodes, the winter/spring ecotype category on this dataset as (along with the 2-row/6-row ecotype) it is the most commonly used physiological means of differentiating barley varieties, and one that all of our test users were familiar with. We also implemented this tool as a web-service which allowed us to include static (but dynamically generated) pedigree representations within our internal barley information portal.

### Feedback on paper-based prototype

Through observation and talking to twelve geneticists and plant breeders while they interacted with our wall-mounted visualization it was clear that there were a number of issues associated with this implementation. Firstly, it was almost impossible to trace edges between nodes when the data was dense (even at a large output size) so we found ourselves falling back on examining text based records to confirm lineage. Secondly, it is incredibly challenging to quickly locate specific plant lines with this density of data. Commonly used lines are immediately identifiable due to the use of size to represent the number of uses in breeding crosses but these are not always what users are most interested in. We also found that users used these larger nodes as reference points, almost as if they were notable points on a map [34, 35] and attempts at using slightly different layouts or orientations were not well received.

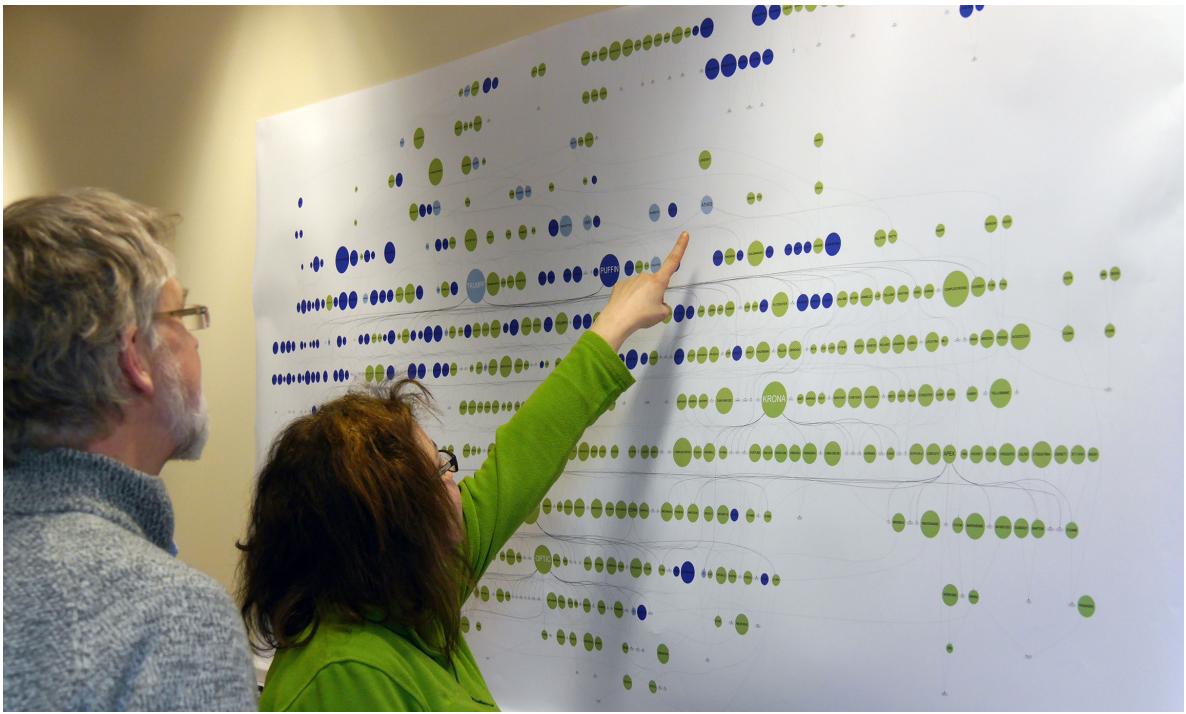

**Figure 6 Helium static prototype.** Users interacting with the large static prototype implementation of our pedigree visualization. In this example 2 and 6 row ecotypes are coloured green and blue respectively and varieties (nodes) are sized based on their contribution to the overall pedigree in terms of the number of lines that are derived from them. This shows which lines are most commonly used in barley breeding in the UK. The 2/6 row division is one of the important characteristics used to differentiate barley worldwide. 2 row barley is primarily of spring type and used in brewing and distilling while 6 row varieties are used in feed due to decreased quality characteristics for the brewing and distilling industry.

It was also clear that users were beginning to quickly spot pedigree problems. These problems related to the parentage of lines and in some cases the assignation of ecotype. These types of errors would be extremely difficult for a user without extensive experience to pick up on and this has not only shown that it is an effective technique for visualization but also an effective way of identifying errors with underlying datasets.

Users liked this representation of large pedigrees. Not only is it visually attractive, but geneticists were using it to identify problems with the underlying pedigree and phenotypic data in a way that is more interactive, social, and tactile compared to the examination of records.

When presented with our results, plant breeders told us that it gave them an overview of their data that was not currently available to them; indeed these representations uncovered interesting information relating to the relative frequency of use of particular "key" lines in the UK Elite Barley germplasm that would have been difficult to see from textual records in the format seen in Figure 2, such records have not been collated like this before. Missing data was also easily spotted thus allowing us to update our underlying datasets.

Problems do however exist, especially in the inability to search for particular plant varieties and tracing of edges to establish lineage. In order to try and address these, it was quickly realised that we would need to move towards the development of a more interactive software tool - Helium - named after the balloon type appearance of our static prototype.

## The Helium prototype

Taking the feedback obtained from our initial informal user testing we implemented an interactive detail and overview [36] prototype pedigree visualization system using Java and the yFiles library from yWorks [37] (Figure 7). We maintained the same visual metaphors (nodes and edges) to describe our pedigree structure but now could add features to allow users to search and explore the data and link in plant passport, phenotype and background data from our Germinate database. One of the design decisions to use Germinate was that we can ensure that researchers working on our barley data will all be using the same data from the same source.

While our paper prototype included a single static image it was clear that when users were viewing our visualization on computer monitors there would be a limitation on the number of nodes that could be displayed while still retaining legibility of line names. To address this our main visualization panel (Figure 7A) can be zoomed and panned to allow users to explore data. We also added an overview panel (Figure 7B) which would allow users to track where they were in the main visualization window and give a high-level overview of the pedigree structure. The overview would act as a common reference point for our users that would not change as the main visualization window was manipulated. Feedback from our paper implementation also showed that users would want to get as much background information as possible on lines and so we added a detail panel (Figure 7C) which displays passport and general background information. Data from Germinate is displayed in the detail panel and is pulled on demand based on a user's selection in the main visualization window.

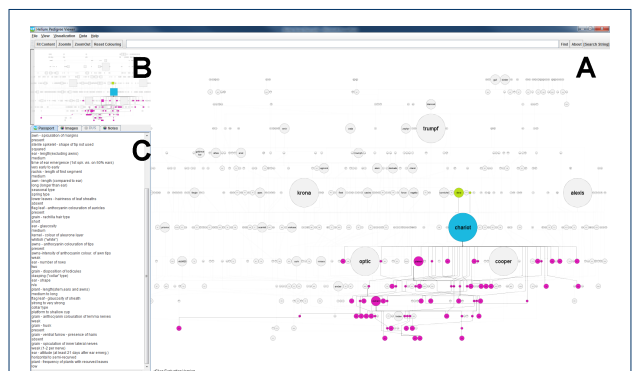

**Figure 7 First Helium Prototype** First Java implementation of Helium showing node sizing and colouring and basic connection to our in-house Germinate barley database. Users can pan and zoom around the display and perform basic searches as well as overlay simple nominal and ordinal data which is loaded from the database backend. This version of Helium was used in subsequent user testing to steer the development of our more advanced system. The colouring in this figure shows predecessors (ancestors) in green, the line of interest in blue and descendants (successors) in purple.

Germinate also includes phenotypic data of both nominal and ordinal types which were colour-coded in Helium using ColorBrewer2 palettes [38, 39]. We use hue to differentiate our nominal data and saturation to distinguish ordinal data classes for phenotypes and saturation to distinguish genetic similarity metrics within our visualization [40] (Figure 8).

## User testing of the Helium prototype

User testing is an important aspect of the development lifecycle of visualization [41, 42, 43]. Both Munzner and Lam lay out the requirements for testing, specifically relating to visualization studies in both contemplation and reflection of user studies.

We performed a subjective evaluation to establish user perception/acceptance and understanding of the visualization methods within Helium. We wanted to establish empirically if users were happy with representing data as graphs, moving away from the traditional family-tree type methods, and whether the use of graphs fits in with a user's perception of pedigree structure and function. Could our users perform basic pedigree operations such as accurately tracking back through generations and find information they require using our visualization? We also wanted to ensure that users were able to interact well with our methods which allow much greater data density and increased plant line density.

The testing data was obtained through a questionnaire and comment-based feedback based on how intuitive our users found the main features of the prototype to be. We also asked how our tool could be improved relating to general usage or new features. This is important as while we had initial user-requirements when our users actually started using our software we had expected them to come up with new ideas on features or utility that would benefit their research.

This feedback allowed us to improve our interface and visualization to help increase our users understanding of the system and underlying biological concepts.

## User testing methodology

We developed a pre-screening questionnaire, user tasks, and a follow up questionnaire centred on predefined tasks that users would be asked to perform. The initial questions were to gain an overall impression of the length of experience the user has had in this field, and to classify their job title. There are two distinct groups of potential users: bioinformaticians/computational biologists and plant geneticists (experimental)/breeders (applied). User tasks were developed using our initial application requirements and were designed to force the users conducting the test to explore our experimental test datasets. The follow up questionnaire was clearly split into two sections; the first taking the form of attitude-scale questions on the user's opinion on the software and visualization in terms of both their use of it (assuming comparison to their current method of viewing these data types), and follow up subjective open-ended questions to get additional information that could be used to drive development of this software tool.

Our questions assume that a comparison is being made to other methods that our test subjects are, or have been using to obtain the same information, and we can use these to signify if our visualization and user interface brings significant improvements in visual representation and understanding of pedigree structure. Throughout the study, notes were taken and screen and audio capture was used to further examine a user's interaction with the interface and to aid in recount of the tests.

Each test was scheduled to take around 45 minutes;

- 5 minutes - pre-questionnaire
- 5 minutes - familiarisation
- 25 minutes - test
- 10 minutes - post-test questionnaire

After completion of the main interaction study our users completed an attitude scale where they indicated their preference on a 5 point scale between "Very Difficult" (1) to "Very Easy" (5) relating to a number of statements about the use of this software.

The questionnaire asked users to detail features or concepts that they found to be confusing, those they found to be clear, and features that they feel would add value to their research. Finally users were asked to provide general comments about their use of our software; we envisage using this to allow us to tweak and fine-tune the Helium interface to aid our users with their research.

### Test results
#### General background profiling
The 16 expert users that undertook this study break down as follows; 5 bioinformaticians, 10 plant geneticists and breeders and 1 statistician. Out of the users 94% were educated to PhD/MSc level and the average length of time working in their areas was 17 years. The minimum experience was 1 year, maximum 36 years giving a median length of experience of 13.5 years.

While all users were familiar with pedigree data, 69% used it on a day-to-day basis as part of their research and 38% regularly used alternative tools.

It should be noted that through verbal feedback it was established that the researchers who were using pedigree data were using paper records and spreadsheets to curate and maintain pedigree data used in their work and not a specific pedigree tool.

#### Main user interaction study
There were eight questions that users were asked to answer in using our pedigree interface. The questions were assigned an overall category and can be seen in Table 1 where we show the question classification along with the number of correct and incorrect responses.

**Table 1** Interaction study correct answers.

| Question classification | Correct (%) | Incorrect |
|---|---|---|
| Unexplained Concepts | 50 | 50 |
| Simple Grandparent Tracking | **93.75** | 6.25 |
| Identifying Children | **56.25** | 43.75 |
| Complex Grandparent Tracking | 50 | 50 |
| Phenotype Classes | **100** | 0 |
| Great-Grandparent Tracking | 37.50 | **62.50** |
| Finding Additional Information | **93.75** | 6.25 |
| Colour Coding Perception | **56.25** | 43.75 |

Our user testing uncovered some interesting problems with our visualization. For example, the category "Identifying Children" from Table 1 asked our participants to identify the progeny of a specific barley variety. In 44% of completed questionnaires this answer was incorrectly given. However, when examining "Tracing Lineage" from Table 2 which related to this question, users thought that it was easy to trace lineage by following graph edges. Our test users were continually missing the same progeny (one of three) of the line; the one whose complete edge was not immediately visible, and disappeared off the right-hand side of their computer display. When talking to a selection of users after the test had been carried out and asking them to perform the same question they did so without error (obviously suspicious to the reasons behind the request).

#### Post-study questionnaires (attitudinal and open ended)
After carrying out our main interaction study the users were asked to fill in a series of questions that asked them to compare Helium to pedigree tools, or methods of handling pedigree data that they are familiar with using, and to get feedback on what they found easy and difficult to understand or perform with Helium. These results are presented in Table 2.

### Test results discussion
We have detailed the most common responses by dividing them into features users liked and disliked. These were from feedback from comments included in our post-study questionnaire.

**Features users liked.** 1. Layout was easy to understand and made scientific sense to users. 2. It was easy to follow edges. 3. Searching for plant lines was simple. 4. Bringing together additional data sources was extremely helpful.

**Table 2** Post Study Questions (Scaled/Likert 5 very easy, 1 very difficult)

| Question Classification | Median (M) | Mode |
|---|---|---|
| Colour Coding | 3 | 3 |
| Phenotype Classes | 4 | 4 |
| Maintaining Position | 4 | 4 |
| Clarity of Relationships | 4 | 5 |
| Tracing Lineage | 4 | 4 |
| Understanding Data | 4 | 4 |
| Background Information | 4.5 | 5 |
| Ease of Use | 4 | 4 |
| Finding Parents | 4.5 | 5 |
| Navigation | 5 | 5 |
| Children | 5 | 5 |
| Finding Lines | 5 | 5 |

***Features users disliked or found confusing.*** 1. Sometimes difficult to differentiate colour coding. 2. Long edges are disorienting. 3. No auto-selection of lines when performing a search. 4. Clearer explanations of ordinal data categories.

Our test users liked the speed at which they could find data, the ease of tracing lineage through complex graphs (although our testing has shown that there were issues with this) and the intuitive layout of our visualization and supporting application. Our testing did highlight some issues, mainly around the use of colour gradients used in ordinal lists which are ineffective and difficult for our users to distinguish when there are more than eight phenotype classes.

## Development of Helium

Feedback from the user evaluation allowed us to address issues that our users had with our prototype in order to develop a more refined and useful visualization application. We needed to work to increase understanding of concepts, representations and visual metaphors that our users found difficult to understand during testing.

The main feedback gained from our initial prototype was that it was difficult to track lineage with overlapping edges and that the ability to interactively overlay, query and retrieve various data types from our internal barley database would be important. Our users also had problems with identifying phenotype classes. Other issues were with the complexity of the graphs and problems identifying children.

Any subsequent development would need to address these points if it was going to offer a usable and effective tool for users.

We re-designed the interface to show 4 main areas: a) the overview panel and data selection panel, b) the main pedigree visualization panel, c) the local view panel and finally d) the details panel. These are described below.

### Overview and Data Selection Panel

This panel (Figure 8A) also includes selection mechanisms for choosing ordinal and nominal categorical phenotypic classes as well as tools for visualizing genetic similarity data (Figure 9). Users can use the overview to navigate to a particular region within the main visualization window if required.

Interactive sliders allow users, in the case of similarity data, to set a percentage similarity value and in real time highlight lines which match the search criteria (Figure 9). In this way it is possible to see lines which should not be closely related appearing on the peripheries of our visualization as the slider is moved, which may indicate problems with pedigree definition or genotyping. We have also included histograms, where appropriate, to show data distribution which can be an aid in the identification of problem markers in which we have addressed in our refined tool by implementing alternative colour schemes. While the number of phenotypes that have this problem is limited, it is nonetheless important to address.

Other features included in this panel are the ability to select more than one phenotype then recolour nodes based on the merged phenotype classes. While originally we had intended to show each phenotype as a different section on a node it was decided, through speaking to users, that they would be interested in finding exact combinations and so we decided to go with the single node colour to reduce clutter and keep the visualization clearer. We do however have problems in the number of colours that may have to be used can be around 20. Such a high number has been shown to be ineffectual at differentiating between classes [44, 40, 45].

### Main Visualization Panel

We modified the main visualization window (Figure 8B) in a number of ways from our prototype. Firstly, we have moved away from bundled orthogonal edge routing (Figure 7) which will make the tracing of lineage easier. We have used slightly modified colour palettes to account for the situation where we have more than eight categorical classes. The new colour palette will help with the problem our testing showed where adjacent classes were too similar in colour for users to accurately distinguish. In Table 6 the incorrect responses to "Identifying Children" were high at 43.75%. In order to address this we have included visual prompts when hovering over a node which will display the number of ingoing and outgoing edges from a node and the names of the line's progeny (Figure 8B). This makes the number of progeny immediately obvious, which we believe will help prevent some of the problems seen in testing. When a user selects a node we have made the edges connecting nodes of interest more prominent by both removing edges which are not associated with the selected node, its ancestors, or successor, and by darkening the edges which are left.

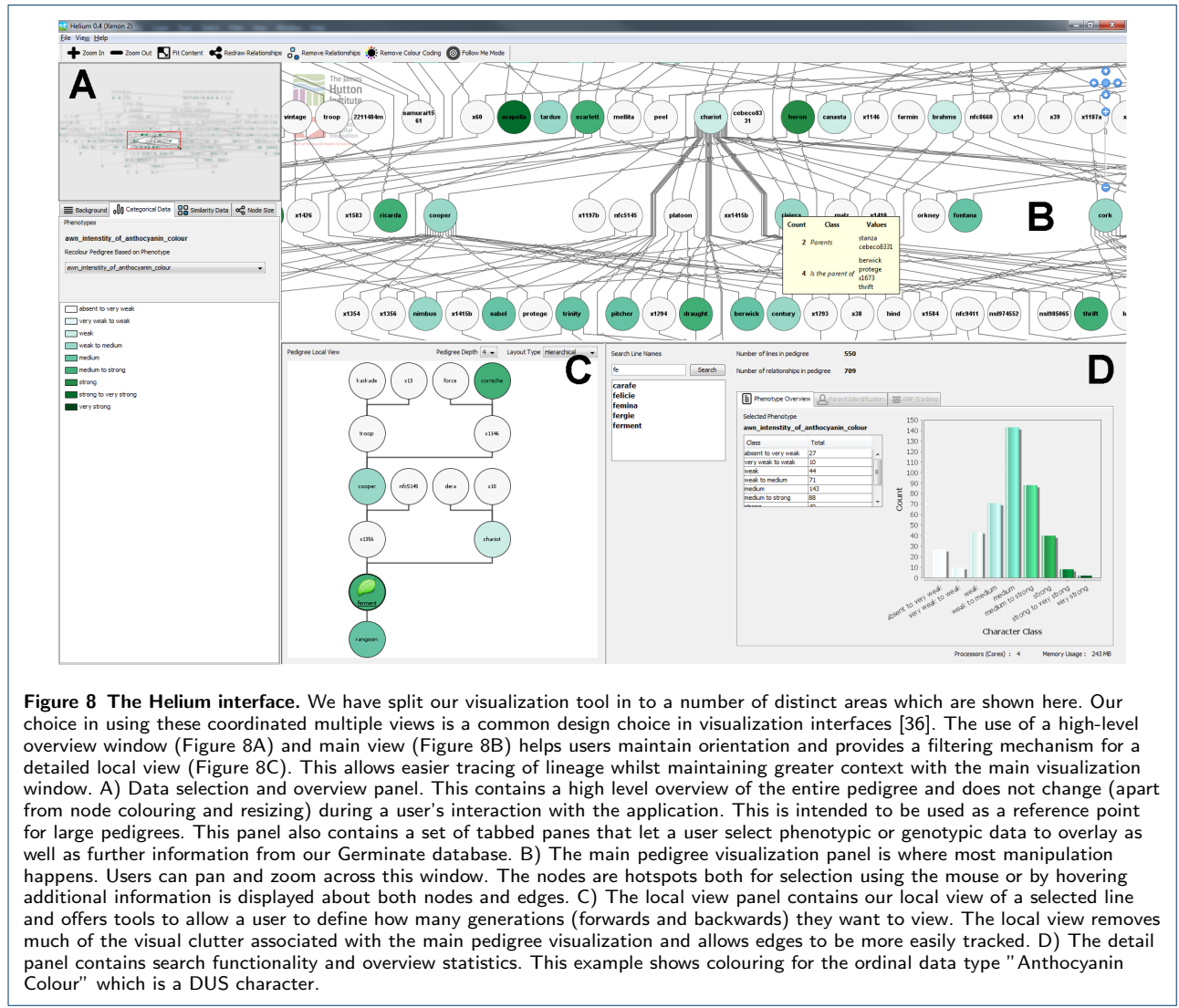

**Figure 8 The Helium interface.** We have split our visualization tool in to a number of distinct areas which are shown here. Our choice in using these coordinated multiple views is a common design choice in visualization interfaces [36]. The use of a high-level overview window (Figure 8A) and main view (Figure 8B) helps users maintain orientation and provides a filtering mechanism for a detailed local view (Figure 8C). This allows easier tracing of lineage whilst maintaining greater context with the main visualization window. A) Data selection and overview panel. This contains a high level overview of the entire pedigree and does not change (apart from node colouring and resizing) during a user's interaction with the application. This is intended to be used as a reference point for large pedigrees. This panel also contains a set of tabbed panes that let a user select phenotypic or genotypic data to overlay as well as further information from our Germinate database. B) The main pedigree visualization panel is where most manipulation happens. Users can pan and zoom across this window. The nodes are hotspots both for selection using the mouse or by hovering additional information is displayed about both nodes and edges. C) The local view panel contains our local view of a selected line and offers tools to allow a user to define how many generations (forwards and backwards) they want to view. The local view removes much of the visual clutter associated with the main pedigree visualization and allows edges to be more easily tracked. D) The detail panel contains search functionality and overview statistics. This example shows colouring for the ordinal data type "Anthocyanin Colour" which is a DUS character.

Hovering over a graph edge will show the names of the two nodes that it connects, in this way with long edges, while using the main visualization window, it is easier to track their origin and destination.

## Local View Panel

Our testing also showed that while users reported they found it easy to identify lineage there were some issues. We felt that these problems could be addressed by including a "local" implementation of our graph showing only the line of interest and its lineage (Figure 8C). This would be shown when a user selects a node in our visualization. We implemented this view below the main visualization window. The local view can be panned and zoomed in the same way as the main visualization window. Within the local view the user has control of how many generations, forwards and backwards, they want to go. This addresses the problems highlighted in Table 1 where there were 50% and 62.5% of users incorrectly answering the "*Complex Grandparent Tracking*" and "*Great-Grandparent Tracking*" questions respectively. With appropriate selection of generation level, grandparents, or indeed any other generation, are now immediately obvious in the simplified pedigree. Additionally we have added the ability to layout the graph using a number of edge routing algorithms. Any changes made to the main pedigree visualization are propagated to the local view. While the local view includes another copy of a portion of the main visualization, it will increase the accuracy of tracing lineage when unnecessary lines are removed and edges between nodes shortened, thus addressing the problems highlighted in testing and reducing the need to "*chase edges*".

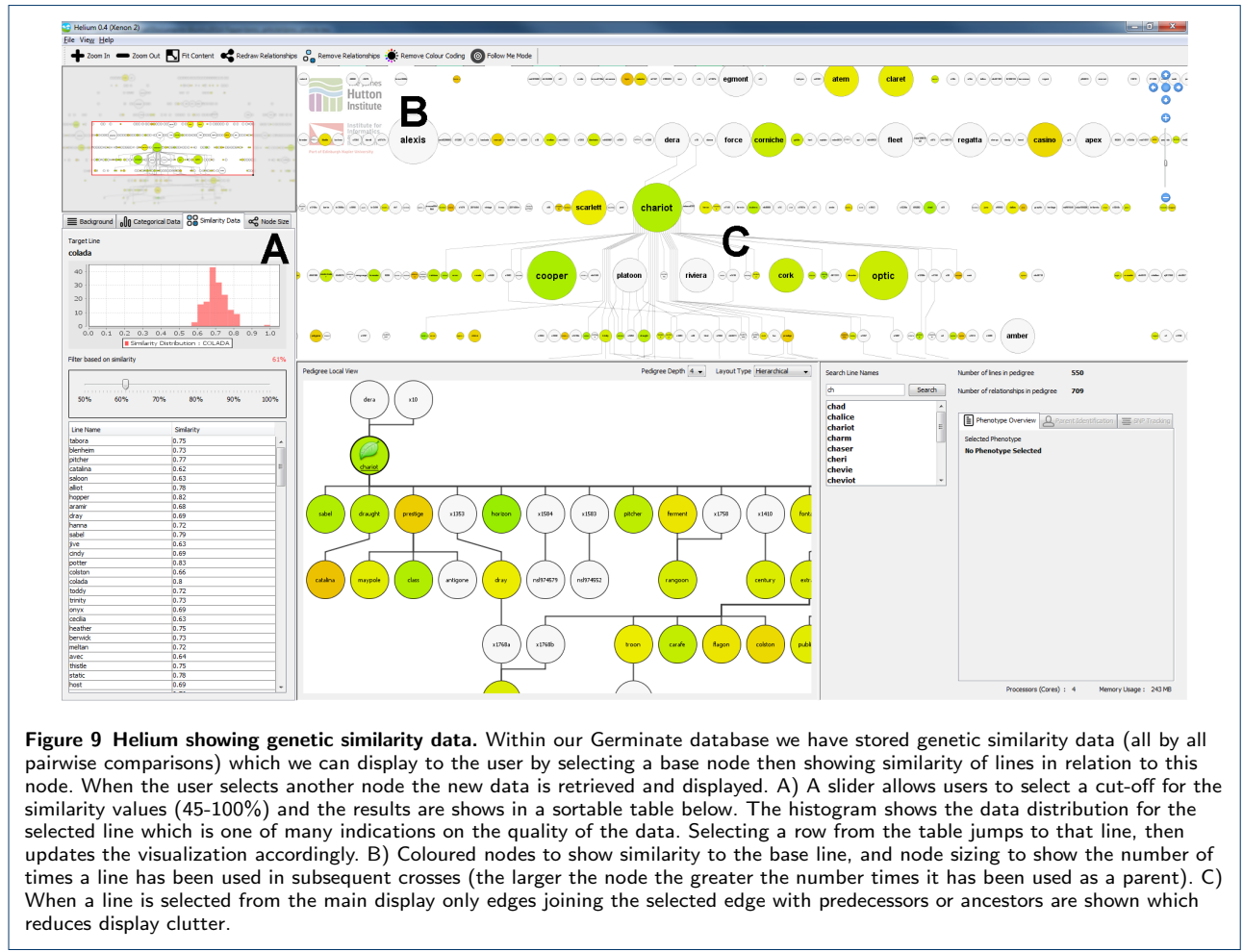

**Figure 9 Helium showing genetic similarity data.** Within our Germinate database we have stored genetic similarity data (all by all pairwise comparisons) which we can display to the user by selecting a base node then showing similarity of lines in relation to this node. When the user selects another node the new data is retrieved and displayed. A) A slider allows users to select a cut-off for the similarity values (45-100%) and the results are shown in a sortable table below. The histogram shows the data distribution for the selected line which is one of many indications on the quality of the data. Selecting a row from the table jumps to that line, then updates the visualization accordingly. B) Coloured nodes to show similarity to the base line, and node sizing to show the number of times a line has been used in subsequent crosses (the larger the node the greater the number times it has been used as a parent). C) When a line is selected from the main display only edges joining the selected edge with predecessors or ancestors are shown which reduces display clutter.

### Detail Panel
The details panel (Figure 8D) shows information about either the current selected phenotype(s) or information from Germinate about specific selected plant lines. In this example we show the distribution of the DUS character "Anthocyanin Colour". We have coloured the histogram in the same way as the phenotype classes in our main visualization window.

The details panel also houses a search functionality which allows searching for lines with usual search features such as wild-card matching and an option which we have called the "follow me" mode which jumps to a search hit, selects it and subsequently updates the detail panel and main visualization window.

During discussions with users it was also apparent that the ability to export line names would be a useful feature to allow scientists to make up lists for sending samples off for genotyping based on phenotypic or genotypic characteristics so we have added functionality to allow users to export lists. Users can select nodes then add them to an export list which can be saved to a text file.

Finally we have included a user history panel which records which lines and phenotypes have been selected over a session so that if required users can go back and see what they had been doing previously. This is important as with large quantities of data it is easy to forget what you have been doing over time.

## Discussion
An interesting outcome of the development of Helium is trying to quantify if what we have developed actually make a user's decision making better and does our tool influence users into making more informed decisions about their data. One of the outcomes from our testing was to assure ourselves that the decisions we had made around the design of the tool were actually good foundations that our target users can build knowledge on and to that end we seem to have made an impact. While we have used standard approaches to the visualization tool we have developed we have applied it directly to a specific domain, and tailored our application appropriately.

While users requested as much information as possible in the interface we need to be careful that we only include necessary information and do not turn Helium into a tool that presents so much unnecessary information to users it in itself becomes unusable or difficult to comprehend; we need to avoid a situation where we overload users with information. While this may seem like a problem that scientists would love to have it could have detrimental effects; do we need to actually present raw data or are overviews enough?. Would a user's understanding be affected by what we present them with?

Users have told us that the overlaying of data onto the pedigree structure has in some ways more impact than showing the division of data in a bar chart or as a table. Having areas of colour in your face brings insight both into the location of clusters of similar data and visual impact of nodes changing from one colour to another, it brings the representation of data to life and in logical an understandable ways.

Examples of the sorts of things that users wanted to be able to do with our tool include a) given genotype data for a line identify possible matches and b) basic error checking based on genotypic or phenotypic data. These are detailed below.

### Given genotype data for a line identify possible matches

Helium will take a string of genotypic data and identify possible matches from data held in our Germinate database then display the possible hits on the pedigree display. This is useful as it is not uncommon for errors to be introduced through mislabelling or handling errors in the lab when genetic material is sent for genotyping. Using the pedigree framework may give users other ways of trying to identify what unknown or problem lines are, or they may point geneticists and breeders in the right direction as to their source, if for example two similar lines are mislabelled we may be able to deduce the correct naming through examination of pedigree records. Further investigation would be required to correctly identify the correct source of this germplasm as there is a possibility either it, or the genotyping is wrong. These types of error are not uncommon.

### Basic error checking based on genotypic or phenotypic data

We can use the interface to look for potential errors with a given line. We know that the alleles of a line must be from either parent, so we can use this in basic error checking. For example, if two lines have been genotyped for allele A at given locus but the progeny has allele B then we know there is a problem. Additionally, we can expand this type of search to look at multiple loci within a dataset. Taking this a step further we can use genotypic data to highlight potential parents of a line and if one parent is known, make a guess at possible candidates for the second parent.

## Conclusions

We have shown through the development of Helium that visualization of our example pedigrees along with genotypic and phenotypic data provides users with new insights into crop breeding.

The representation of our unique barley test dataset shows that the pedigree structure takes the form of what we have coined a pedigree net. Our visualization has shown that there are three main classes of plant lines seen when we view them in Helium which we have named; a) principal lines which are commonly used to generate new cultivars due to their possession of desirable characteristics b) flanking cultivars brought in to increase the genetic diversity of subsequent lines and less commonly used in crosses and finally c) terminal varieties that are released, but have had little subsequent use.

One of the more hard-hitting measures of success of our first paper-based prototype came from the presentation of data to a meeting of UK plant breeders. While the pedigree data that we demonstrated was available to all in the room as written records, (like those in Figure 2), the representation that we showed (Figure 5) had a major impact through the provision of new insights as to how germplasm was very closely related. When written as a text string it is difficult to construct the bigger picture, but when displayed in our tool, the relationships between competing breeders lines was much more striking. While this was privately known to the individual breeders, having it presented to them when they were all in the same room was very enlightening. This not only highlights the value of visualization but that we have implemented a visualization tool with real-world impact.

While we have tailored Helium to specific data types(genotypic/similarity, nominal and ordinal phenotypic data and pedigree definitions) we intend it to be a framework on to which, over time, additional data types can be added and we are working with worldwide plant scientists and breeders to develop the Helium platform further.

For more information on Helium please visit our website http://ics.hutton.ac.uk/helium.

**Authors' contributions**
PDS wrote this paper, developed the pedigree visualizations and database infrastructure, and designed and conducted the user study. IM gave advice on programming and computing requirements. MG, JK and DFM contributed towards the writing and editing of the manuscript and provided constant advice over the course of this work.

**Acknowledgements**
The authors gratefully acknowledge funding from the Scottish Government's Rural and Environment Science and Analytical Services (RESAS) division and Edinburgh Napier University. We would also like to thank colleagues at The James Hutton Institute, in particular Bill Thomas and Luke Ramsay for help and advice with pedigree data. We would also like to thank colleagues from NIAB (National Institute of Agricultural Botany) and the AGOUEB (Association Genetics of UK Elite Barley) consortium for the use of experimental data. Additionally, we would like to thank those who were generous enough with their time and enthusiasm to participate in the user evaluation of this software tool.

**Author details**
[1]Information and Computational Sciences, The James Hutton Institute, Invergowrie, DD2 5DA Dundee, UK. [2]School of Computing, Edinburgh Napier University, 10 Colinton Road, EH10 5DT Edinburgh, UK.